\documentclass[amsmath,amssymb,aps,eqsecnum,twocolumn,nofootinbib]{revtex4}
\usepackage[dvips]{graphics,color}
\usepackage{latexsym}
\usepackage{graphicx}
\begin{document}


\title{Bubbles in the Self-Accelerating Universe}

\author{
Keisuke Izumi$^{a}$\footnote{e-mail:
ksuke@tap.scphys.kyoto-u.ac.jp}, Kazuya
Koyama$^{b}$\footnote{e-mail: kazuya.koyama@port.ac.uk}, Oriol
Pujol{\`a}s$^{c}$\footnote{e-mail: pujolas@ccpp.nyu.edu} and
Takahiro Tanaka$^a$\footnote{e-mail:
tama@tap.scphys.kyoto-u.ac.jp}\\~}

\address{$^a$Department of Physics, Kyoto University, Kyoto 606-8502, Japan}
\address{$^b$Institute of Cosmology and Gravitation, University of
Portsmouth, Portsmouth PO1 2EG,UK}
\address{$^c$Center for Cosmology and Particle Physics,
New York University New York, NY, 10003, USA}

\begin{abstract}

We revisit the issue of the stability in the
Dvali-Gabadadze-Porrati model, by considering the nucleation of
bubbles of the conventional branch within the self-accelerating
branch. We construct an instanton describing this process in the
thin wall approximation. On one side of the bubble wall, the bulk
consists of the exterior of the brane while on the other side it
is the interior. The solution requires the presence of a 2-brane
(the bubble wall) which induces the transition. However, we show
that this instanton cannot be realized as the thin wall limit of
any smooth solution. Once the bubble thickness is resolved, the
equations of motion do not allow $O(4)$ symmetric solutions
joining the two branches. We conclude  that the thin wall
instanton is unphysical, and that one cannot have processes
connecting the two branches, unless negative tension bubble walls
are introduced. This also suggests that the self-accelerating
branch does not decay into the conventional branch nucleating
bubbles. We comment on other kinds of bubbles that could
interpolate between the two branches.
\end{abstract}

\maketitle

\section{Introduction}

The acceleration of the universe is one of the most important
problems in modern cosmology. An interesting approach to explain
it is to modify gravity at very large distances. One of the most
studied examples of a modified gravity model is the
Dvali-Gabadadze-Porrati (DGP) brane-world model~\cite{DGP}. In
this model, there exits a self-accelerating (SA) solution in which
the universe undergoes an exponential expansion without
introducing a cosmological constant~\cite{cosmology}. However,
this branch contains a ghost in the spectrum at the quadratic
order~\cite{effective1, effective2, Koyama1, Koyama2,
Charmousis,Carena,Izumi}. It has been argued~\cite{dgi,dvali06}
that a perturbative argument is not conclusive because the brane
is in a strongly coupled regime in this branch, and it is not
clear that this implies an instability (see also the recent
analysis of nonlinearities in a cosmological background given in
Ref.~\cite{koyamaSilva}).

There is some indication that the SA branch is pathological at the
nonperturbative level~\cite{Gabadadze06}, coming from the known
exact solutions for domain walls~\cite{domain} and point-like
sources~\cite{GI} localized on the brane (although the bulk
solution is not known in the latter case). In both cases, the
effective gravitational mass (or tension) of the source is
negative from the five dimensional point of view, which might
signal an instability in the full nonlinear theory. This raises a
number of interesting questions. If the SA branch contains a ghost
instability at nonperturbative level, then we are naturally lead
to ask what does this solution decay to. Is it possible to have
transitions from one branch to the other? This is what we shall
address in this paper, with the hope that it can shed some light
on the presence of ghosts in the SA solution and the question to
what extent they are harmful.

The 5D action describing the DGP model is given by
\begin{equation*}
S = \int d^5 x \sqrt{-g} \, \frac{{}^{(5)\!} R}{2 \kappa^2}  +
\int d^4 x \sqrt{-\gamma} \, \left( \frac{R}{2 \kappa_4^2}
+\frac{K^\mu_{~\mu}}{\kappa^2} -\sigma\right) , \label{action}
\end{equation*}
where $\sigma$ is the brane tension and $K^{\mu}_{\:\: \mu}$ is
the trace of the extrinsic curvature. We assume $Z_2$ symmetry
across the brane, and above we omitted the matter Lagrangian. This
model has two very interesting features. The first is that gravity
becomes five dimensional at distances larger than the crossover
scale
\begin{equation}
r_c = \frac{\kappa^2}{2 \kappa_4^2}.
\end{equation}
The other is that for generic sources, it allows two branches. For
example, considering only the brane tension $\sigma$, the flat FRW
solutions can take two different values of Hubble parameter on the
brane given by~\cite{cosmology}
\begin{equation}
H_\pm = |K_\pm|,
\end{equation}
with
\begin{equation}
K_\pm = \frac{1\pm \sqrt{1+ 4 \,r_c^2 \kappa_4^2\, \sigma
/3}}{2r_c}~, \label{Hpm}
\end{equation}
where $K_\pm$ is quarter the trace of the extrinsic curvature of
the brane in the respective branches.
The difference between the two branches can be visualized (for
$\sigma>0$) as the different embeddings of the brane. In both
solutions, the brane is a 4D de Sitter hyperboloid in the 5D
Minkowski spacetime. For the conventional (self-accelerating) branch,
the bulk is identified as the interior (exterior) of the
hyperboloid. The SA solution is the one with the larger
Hubble rate, $H_+$. It is phenomenologically interesting that
the Hubble parameter at late times in the SA branch
approaches a constant $H = 1/r_c$ even for vanishing $\sigma$,
mimicking a cosmological constant~\cite{cosmology}.

As has been pointed out in Refs.~\cite{effective1, effective2,
Koyama1, Koyama2, Charmousis, Carena}, the SA universe contains a
ghost. For a positive tension ($Hr_c>1$), one has  a spin-2
graviton having mass in the range $0 < m^2 < 2H^2$ and its
helicity-0 excitation is a ghost~\cite{Koyama1}. On the other
hand, for a negative tension brane ($Hr_c <1$), the spin-0 mode
becomes a ghost. In the SA universe ($Hr_c=1$), the spin-0 mode
and a helicity-0 excitation of the spin-2 mode mix and a ghost
arises from the mixing~\cite{Koyama2}. It was shown that the ghost
cannot be removed even in the presence of the second brane in the
bulk due to the correlation between spin-0 and spin-2
ghosts~\cite{Izumi}.
On the other hand, the conventional
branch of solutions does not contain
ghosts, though it does not self-inflate.

Having a ghost in the spectrum for the SA universe,
one expects that it is unstable to spontaneous particle creation
out of the vacuum or to some even more dramatic process. In this
article, we try to go one step further in the  analysis of the
stability of the SA branch by considering nonperturbative decay
channels. Since the conventional branch solution is stable, it is
tempting to think that the SA solution decays into
the conventional branch solution. Here we shall consider the nucleation
of bubbles of the conventional branch in the environment of the
SA branch. This would resemble a kind of false
vacuum decay in de Sitter space.
False vacuum decay is described
by an instanton which is a classical solution in an Euclidean time
connecting initial and final configurations. In our case, we are
interested in a solution that interpolates between the
SA and the conventional branches~\cite{Tanakainst}.

We shall see that in the thin wall approximation, there exists a
solution similar to the Coleman de Luccia instanton~\cite{cdl}
describing this process.
Unlike the bubble of nothing~\cite{nothing}, mediating the decay
of the Kaluza-Klein vacuum, the bubbles do not occur spontaneously
in the gravitational sector in the DGP model. Rather, a material
wall separating the two phases is required for the instanton to
exist.

The second part of this paper is dedicated to show that this
solution cannot be recovered as a limit of any microscopic model
of the bubble based on a scalar field. We interpret this as an
indication that the thin wall instanton is unphysical. Hence, the
SA branch does not decay nucleating bubbles of the
conventional branch. In Section \ref{sec:disc}, we comment on other
possible decay channels.

\section{Instanton in the thin wall approximation}

Let us construct the instanton representing a bubble of the
conventional branch in the background of the SA
branch. We follow the method of Refs.~\cite{PG}, which was
recently applied to the analysis of domain wall in the DGP
model~\cite{domain}. However, our result is different from
Refs.~\cite{PG} in the calculation of the matching condition on
the bubble wall, as was also pointed out by Ref.~\cite{domain}.

The bulk spacetime is the 5D Minkowski spacetime. In Euclidean
time, the bulk metric is
\begin{equation}
 ds^2=dz^2+dr^2+r^2 d\Omega^2,
\end{equation}
where $d\Omega^2$ denotes the line element on a unit 3-sphere. The
trajectory of the brane in this 5D Euclidean space is specified by
two functions $(z(\tau),r(\tau))$, and we can always choose the
gauge condition
\begin{equation}\label{gauge}
  \dot z^2+\dot r^2=1,
\end{equation}
where a dot ``$~\dot{}~$'' represents differentiation with respect to $\tau$.
Then, the induced metric on the brane is given by
\begin{equation}
ds_4^2 = d \tau^2 + r^2(\tau) d\Omega^2.
\end{equation}
The equations that determine the embedding are Eq.~(\ref{gauge})
and the Israel junction conditions
\begin{equation}\label{israel}
K^\mu_\nu - \delta^{\mu}_{\nu} K = - \frac{\kappa^2}{2}
\left(T^{\mu}_{\nu} - \frac{1}{\kappa_4^2} G^{\mu}_{\nu} \right).
\end{equation}
The $(\tau, \tau)$-component of this equation becomes
\begin{equation}\label{tautau}
6 \frac{\dot z }{r}=\kappa^2 \left( \sigma- \frac{3}{\kappa_4^2}
\frac{\dot z^2}{r^2} \right).
\end{equation}
The solution to Eq.~(\ref{gauge}) and Eq.~(\ref{tautau}) is
\begin{eqnarray}
r_{\pm}(\tau)&=& 
K_{\pm}^{-1}\sin \big[K_{\pm}
(\tau^0_{\pm}+\tau) \big] , \cr%
z_{\pm}(\tau)&=& 
K_{\pm}^{-1}\cos \big[K_{\pm} (\tau^0_{\pm}+\tau) \big] +
z_\pm^0~, \label{sol}
\end{eqnarray}
where the $+$ ($-$) subscript refers to the form of the solution
in the self-accelerating (conventional)
branch segment, and holds for $\tau>0$
($\tau<0$).
When the extrinsic curvature is positive (negative), the
exterior (interior) of the sphere is the bulk.
Therefore, when $\dot{z}>0$ ($\dot{z}<0$),
the bulk is the region where the value of $r$ is
smaller (larger)
than the one on the brane.
With this choice,
the bulk is guaranteed to be consistently
connected on the same side
of the brane at the interface of the different segments.
One of the
constants $\tau^0_\pm$ is fixed by the junction condition on the
bubble wall (see below) and the other by continuity. Similarly,
one of the $z^0_\pm$ is fixed by continuity and the other can be
set to zero with no loss of generality. Fig.~1 shows the behaviour
of these solutions in $(z,r)$ plane.

\begin{figure}[h]
\label{thinbubble}
 \begin{center}
\includegraphics[width=8cm]{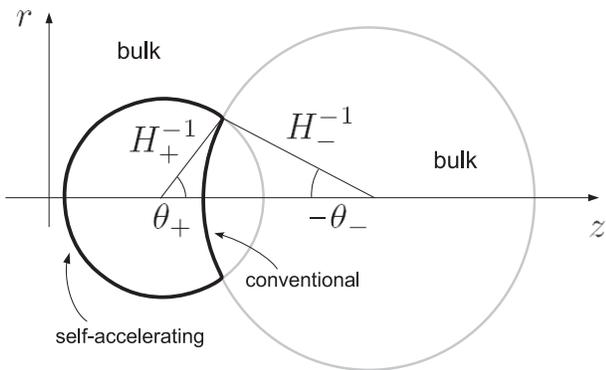}
\caption[]{Trajectory of the brane in $(z,r)$ plane (for
$\sigma>0$).}
\end{center}
\end{figure}

The continuity of $r$ at $\tau=0$ requires
\begin{equation}
\frac{\sin\theta_+}{K_+}  = \frac{\sin\theta_-}{K_-} ,
\end{equation}
where $\theta_\pm =K_\pm \tau^0_\pm$.
The angular components of the junction condition
(\ref{israel}) have a delta function singularity at $\tau=0$ as
$\dot{r}$ is discontinuous. The singular part of the junction
condition close to $\tau=0$ is given by
\begin{equation}
2 \frac{\ddot{r}}{\dot z} = -\kappa^2
\left( \mu \delta(\tau) + \frac{2}{\kappa_4^2} \frac{\ddot{r}}{r} \right).
\label{singular}
\end{equation}
The term on the left-hand side of the above equation
is simply written as $2\dot\theta$
by introducing the parametrization $\dot r=\cos\theta$
and $\dot z=-\sin\theta$.
Integrating over the infinitesimal interval around $\tau=0$,
and using Eq.~(\ref{sol}), we obtain
the following jump condition on the bubble
\begin{equation}
2(\theta_+ - \theta_-)+4r_c\left(K_+ \cot\theta_+ - K_- \cot\theta_-
\right)=- \kappa^2\mu,
\end{equation}
where $\mu$ is a tension of the bubble wall. The first term is the
deficit angle arising from the 5D Einstein tensor and the second
term is the jump of the 3D extrinsic curvature across the bubble
wall coming from the 4D Einstein tensor.\footnote{Note that in
Refs.~\cite{PG}, the integration of the left-hand side of
Eq.~(\ref{singular}) was done
with $\dot z$ fixed. This is  not true
and thus our result for
the 5D singular part is different from Refs.~\cite{PG} (see also
Ref.~\cite{domain}).}

The second term on the left-hand side
can be rewritten as
\begin{eqnarray}
&& r_c\left(K_+ \cot\theta_+  - K_- \cot\theta_- \right)\nonumber\\
&&\qquad =  -r_c{K_i\over \sin\theta_i}\left(\sin\theta_+
+\sin\theta_-\right)
\tan{\theta_+ - \theta_-\over 2} \nonumber\\
&&\qquad  =  - \tan{\theta_+ - \theta_-\over 2}~,
\end{eqnarray}
where we used the identity
\begin{equation}
{\cos\theta_+ - \cos\theta_- \over \sin\theta_+ + \sin\theta_-}
 = -\tan {\theta_+ - \theta_- \over 2},  \nonumber
\label{key1}
\end{equation}
and $r_c(K_+ + K_-)=1$, which immediately follows from
Eq.~(\ref{Hpm}). 
Thus we obtain an equation for $\overline\theta \equiv
(\theta_+ - \theta_-)/2$,
\begin{equation}
(\tan{\overline\theta})
-\overline\theta ={\kappa^2 \mu \over 4}, \label{instantoneq}
\end{equation}
which directly determines the deficit angle, $-4\overline\theta$.
If there is no domain wall tension, Eq.~(\ref{instantoneq})
immediately implies that there is no instanton that describes the
transition from the SA universe to the conventional
branch solutions. Thus the SA universe cannot decay
into the conventional branch solution without introducing a domain wall
tension. On the other hand, Eq.~(\ref{instantoneq}) admits a
solution if we have the positive tension wall $\mu >0$. This would
indicate that a 2-brane can induce the SA universe
to decay into the conventional branch.

When both $\kappa^2\mu$ and $\kappa_4^{2}r_c^{2}\sigma$ are small,
$|\theta_-|\ll\theta_+\ll 1$ holds. In this case, we have
\begin{eqnarray*}\label{deficit}
\theta_+\simeq \left(6\kappa^2\mu\right)^{1/3}.
\end{eqnarray*}
The difference between the instanton and the background action is
\begin{equation}
\Delta I \approx 12\pi^2{\theta_+^4 r_c^3\over \kappa^2}.
\end{equation}
The suppression of the rate of this tunnelling process is of order
$\sim e^{-\Delta I}$. Thus, bubbles with a low enough
tension, as given by
$$
\kappa^{2} \,\mu  \ll ( r_c/\kappa_4)^{-3/2} ~,
$$
have an unsuppressed tunnelling rate. For $r_c$ of order of the
present Hubble radius we have $r_c/\kappa_4\approx 10^{60}$, which
leads to a tension of order $10^{-90} \kappa^{-2}$ or less.

\section{Non-existence of smooth branch-changing solutions}

Let us now show that any attempt to find a smooth version of the
instanton of the previous section which interpolates between the
two branches is doomed to fail. We shall assume that the bubble is
supported by a scalar field $\phi$ localized on the brane, with
some potential $V(\phi)$ which we do not need to specify.
Generalizing a constant $\sigma$ to a function $\rho_E$,
Eqs.~(\ref{gauge}) and (\ref{tautau}) are more conveniently
written in this case as
\begin{equation}\label{zdot}
-\frac{\dot z }{r}=\hat K_\pm(\rho_E)~,
\end{equation}
and
\begin{equation}\label{r}
 \frac{1-\dot r^2}{r^2}=\hat K_\pm^2(\rho_E)~,
\end{equation}
where
\begin{equation}\label{HRhoE}
\hat K_\pm(\rho_E) \equiv \frac{1 \pm \sqrt{1+ 4 \,r_c^2 \kappa_4^2\,
\rho_E /3}}{2 r_c} ~,
\end{equation}
and the ``Euclidean energy density'' is given by \footnote{ It is
manifest that the positivity of $\rho_E$ is not guaranteed in
general even for a normal field as is clear from Eq.~\eqref{rhoE}.
Here we used the notation $\rho_E$ from the analogy to the closed
FRW model, but its analytic continuation to the Lorentzian region
does not mean the energy density in general. When the expression
is analytically continued through the $\tau=$ constant maximal
hypersurface given by $\dot r(\tau)=0$, the analytic continuation
of $\rho_E$ becomes the energy density. In the context of the
bubble nucleation, however, the hypersurface corresponding to the
junction between Euclidean and Lorentzian regions is not given by
a $\tau=$constant surface. In this case, the analytic continuation
of $\rho_E$ becomes minus the pressure along the direction
orthogonal to the bubble wall, which is not directly related to
the wall tension. }
\begin{equation}\label{rhoE}
\rho_E = - \frac{1}{2} \dot{\phi}^2 + V(\phi).
\end{equation}
We have chosen ``$-$'' sign on the left-hand side in
Eq.~(\ref{zdot}). This sign is conventional, but once we choose it
we have to keep it fixed during the calculation. It is clear from
(\ref{HRhoE}) that a solution that smoothly connects the two
branches at some point must have $\rho=-\rho_c$, where
\begin{equation}
  \rho_c\equiv \frac{3}{4 \, \kappa_4^2 r_c^2}~.
\end{equation}
Even if we allow for negative value of $\rho_E$, we can easily
show that there are no branch-changing solutions.
In order to interpolate between the two branches, $\rho_E$ has to
reach the value $-\rho_c$, as shown in Fig.~2. Hence, $\dot
\rho_E$ must change sign at the point where $\rho = -\rho_c$. On
the other hand, the equation of motion for $\phi$ is equivalent
(as long as $\dot\phi\neq0$) to
\begin{equation}\label{conservation}
\dot{\rho}_E = 3 \frac{\dot{r}}{r} \dot{\phi}^2~.
\end{equation}
This means that $\dot \rho_E$ can only change sign if $\dot r$
also does, and consequently that if $\rho_E$ develops a minimum,
then $r(\tau)$ also does. Given that the minimum of $\rho_E$ must
be at $-\rho_c$, Eqs.~(\ref{r}) and (\ref{HRhoE}) imply that the
minimum in $r(\tau)$ occurs
with $r= 2 r_c$,
as illustrated in Fig.~3. The key point is that having a minimum at
this value is incompatible with Eq.~(\ref{r}), when we try to
integrate it with the choice of sign corresponding to the SA
branch. Indeed, if $r(\tau)$ has a minimum at $2r_c$, then the left-hand side
of Eq.~(\ref{r}) is bounded above by $1/(2r_c)^2$. But in the
SA branch we have $\hat K_+^2>1/(2r_c)^2$.
This is a contradiction.
Note that this proof does not rely on any assumptions of the
regularity of the instanton ({\em e.g.}, at the poles, $r=0$), so
even if singular solutions are allowed in the path integral, they
would not include an $O(4)$ symmetric branch-changing
configuration.
Hence, we
conclude that there is no smooth solution connecting
two different branches.

\begin{figure}[t]
  \begin{center}
    \includegraphics[height=5.5cm]{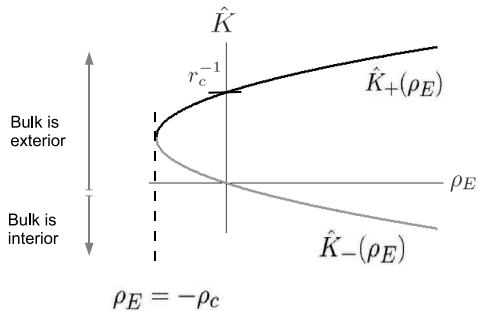}
  \end{center}
\vspace*{-1cm}
  \caption{Relation between $\rho_E$ and $\hat K$.
            we do not have to consider dashed line. Any solution
            that smoothly joins the two branches must have
            an energy density $\rho_E=-\rho_c$ at some point.}
  \label{fig:K.eps}
\end{figure}

\begin{figure}[t]
\label{thickbubble}
 \begin{center}
\includegraphics[keepaspectratio=true,width=8cm]{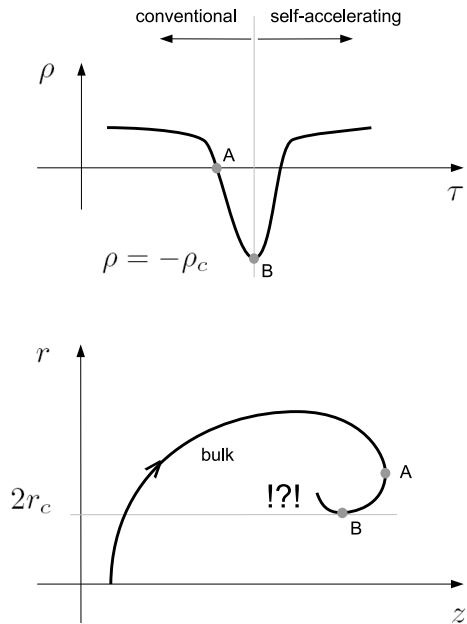}
\vspace*{-1cm} \caption[]{In the upper panel we sketch the form of
$\rho_E(\tau)$ for a solution that smoothly joins the two
branches. At the point $B$, we have $\rho=-\rho_c$. In the lower
panel, we show schematically the trajectory starting from the
conventional branch. At point $A$, $\rho_E$ vanishes, and so does
the extrinsic curvature. The bulk is the interior ($r<r(\tau)$)
before that point and the exterior ($r>r(\tau)$) after it.  At
point $B$, by the conservation equation, $r(\tau)$ develops a
minimum at $2r_c$. However, matching to the SA branch at point $B$
is incompatible with Eq.~(\ref{r}).}
\end{center}
\end{figure}

By contrast, if and only if we stay on the conventional branch on
both sides of the point at which $\rho=-\rho_c$, the equations for
the brane trajectory can be integrated around the minimum. This is
the only possible configuration compatible with the profile of
$\rho(\tau)$ shown in upper panel of Fig.~3.

In summary, the argument for the absence of branch-changing
solutions consists of two observations. The first is that, in
order
to join the two branches, 
$\rho_E(\tau)$ must have a minimum at $-\rho_c$.
The second is that a minimum in $\rho_E(\tau)$ is only compatible
with the equations of motion if we stay on the conventional branch
around the minimum.

It is instructive to compare this situation to what we would find
in GR under the assumption that $\rho_E(\tau)$ has a minimum. By
the conservation equation (\ref{conservation}), one also concludes
that $r(\tau)$ has a minimum. In GR, we would not have Eq
(\ref{zdot}), and $\hat K_\pm^2$ in Eq.~(\ref{r}) would be
replaced by $\kappa_4^2\rho_E/3$. Since this is always increasing
in $\rho_E$, it follows by the same argument that a minimum in
$r(\tau)$ would be incompatible with Eq.~(\ref{r}) (irrespective
of the value of $\rho_E$ at the minimum). Hence, the
reason why one can obtain a minimum in DGP for the conventional
branch relies on the very peculiar property that
$$
{\partial \hat K_-^2(\rho_E)\over \partial\rho_E} < 0
$$
in the range $-\rho_c<\rho_E<0$. The Hubble rate in the closed FRW
universe is given by $H^2=\hat K_-^2(\rho)$ where $\rho$ is the
energy density. Hence, in this range $H^2$ increases with
decreasing the energy density $\rho$, which is a manifestation of
anti-gravity. This is expected since for very low densities
gravity in DGP behaves as in 5D, where the sign of the effective
Newton constant is reversed for $\rho_E<0$.

The obstruction to find a minimum arises because Eq.~\eqref{rhoE}
enforces that the signs of $\dot r$ and $\dot\rho_E$ are
identical. Generally, this relation holds if $\rho_E+P_E<0$, where
$P_E$ is the analytic continuation of the pressure to Euclidean.
The energy momentum tensor on the four dimensional brane takes the
form, $T^\mu_\nu={\rm diag}(-\rho_E,P_E,P_E,P_E)$. In the case of
a scalar field, we have $P_E=-\dot\phi^2/2-V(\phi)$. As $P_E$
dominates in the thin wall case, the above condition
$\rho_E+P_E<0$ is approximately interpreted as the negative
pressure of the bubble wall, which means that the tension of the
bubble wall is positive. Then, permitting the presence of a
negative tension bubble drastically alters our discussion. In
particular, it is not difficult to construct branch-changing
instantons with a pair of positive and negative tension bubbles.
However, a normal scalar field cannot support a negative tension
bubble. The only way to incorporate a negative tension bubble
using a scalar field model is to flip the overall sign of its
action, \emph{i.e.} to introduce a ghost. In general, it seems
that the negative tension bubble requires the presence of an
additional ghost matter sector\footnote{This ghost should be
distinguished from the familiar ghost which resides in the
gravitational sector in the SA branch. The gravitational ghost is
built into the DGP model from the beginning, and it is already
taken into account in our present analysis.}. Therefore we shall
not pursue this direction here. (See Ref.~\cite{eg} for a recent
discussion on the nucleation of pairs of ghost and normal domain
walls.)

\section{Discussion}
\label{sec:disc}

In summary, we have studied the possibility to construct an $O(4)$
invariant instanton that describes the nucleation of bubbles of
the conventional branch in the self-accelerating (SA) universe. We found that
this is impossible unless there is a material bubble wall which
induces the transition. Introducing the bubble wall tension, one
can find an instanton in the thin wall limit. However, we showed
that it is impossible to construct this instanton in a scalar
field model of the wall.
The reason is that at the point
where the two branches are connected, the equations of
motion cannot be integrated in the SA branch. It immediately
follows that the thin wall instanton is unphysical
and that
transitions between the two branches cannot happen  (at least with
$O(4)$ symmetry).

We should emphasize that the obstruction to have the solution is
not due to the impossibility to embed the brane in such a way that
the bulk is the exterior of the brane on one side of the wall and
the interior on the other\footnote{To illustrate this point,
consider a potential $V(\phi)$ with a `false' vacuum with positive
cosmological constant (CC) and a `true' vacuum with negative CC
(but larger than $-\rho_c$). Then, true vacuum bubbles solutions
within the same branch can form. For the conventional branch, the
instanton looks like Fig.~1 with the bulk being the complement of
what is shown there, so that the bulk corresponds the interior or
the exterior of the brane in either side of the bubble wall.}.
Instead, it is a direct consequence of the equivalent of the null
energy condition ($\rho_E+P_E<0$), which is satisfied in a scalar
model of the bubble. Let us now give a few comments on the
interpretation of our results.\\

\emph{Bubbles in the `$\pi$-Lagrangian'}\\
It is also useful to analyze the bubble configurations in terms of
the effective scalar field Lagrangian introduced in
Ref.~\cite{lpr} (see also Ref.~\cite{nr} and \cite{GIdec}) to
describe the decoupling limit. In this limit, the most relevant
non-linearities are encoded in the brane bending mode $\pi$, and
are described by the equation of motion
\begin{equation}\label{pi}
3 \Box \pi + \frac{(\Box \pi)^2 - (\nabla_{\mu} \nabla_{\nu}
\pi)^2}{\Lambda^3}=-\kappa_4 \;{T\over 2}~,
\end{equation}
where $\Lambda^3 =(2\kappa_4 r_c^{2})^{-1}$ and $T$ is the trace
of the stress tensor of  $\phi$. In this treatment, the equation
of motion for $\phi$ is the same as in flat space. In the absence
of the brane tension, the conventional branch is $\pi_0=const$ and
the SA branch is $\pi_0 = -(\Lambda^3/2)\, x^\mu x_\mu $. One
expects that analog of the thin wall instanton of Sec.~II is
\begin{equation}\label{piguess}
\pi=-(\Lambda^3/2)\;(\xi^2-\xi_0^2)\;\Theta(\xi^2-\xi_0^2)~,
\end{equation} %
where we introduced the `Rindler' coordinate
$\xi=(x^\mu x_\mu)^{1/2}$. Inside the hyperboloid
at $\xi=\xi_0$ there is the conventional
branch and outside the SA branch. Clearly, the only nontrivial
step to check whether this is a solution of Eq.~\eqref{pi} is to see what
happens on the hyperboloid. Assuming that $\pi$ depends only on $\xi$,
the left-hand side
of Eq.~\eqref{pi} is rewritten as 
$$
\left(1+{2\over \Lambda^3}{\pi'\over \xi}\right)\pi''
+\left(3+{2\over \Lambda^3}{\pi'\over \xi}\right)\;{\pi'\over
\xi}, 
$$
where $\pi'=\partial_\xi \pi$. Hence, the singular term at $\xi=\xi_0$
on the left-hand side has a distributional form like ${\rm
sign}(\xi-\xi_0)\,\delta(\xi-\xi_0)$ (and it is proportional to $\xi_0$).
This means that the configuration
\eqref{piguess} cannot be sourced by a
simple $\delta$-function.
In fact, integrating around the bubble wall, one obtains
$$
\Delta \pi' + {1\over \Lambda^3 \xi_0} \Delta (\pi')^2 ,
$$
where $\Delta \pi' =
\pi'|_{\rm SA}-\pi'|_{\rm conventional}$. Given that
 $\pi'=-\Lambda^3 \xi$ in the
SA branch, this is identically zero! This does not mean that this
is a `vacuum' (no source) solution, but rather that the total
tension of the source vanishes, as would happen for example in a
system of two walls with opposite tensions. The fact that the two
terms in the previous equation cancel out is the remnant of the
cancellation between the terms linear in $\overline\theta$ on the
left-hand side of Eq.~\eqref{instantoneq}. Recall that these terms
originate from the intrinsic curvature and the extrinsic
curvature, respectively. In the exact solution
\eqref{instantoneq}, these terms do not cancel out at higher order
in $\overline\theta$, and this is what allows to have solutions
with a net bubble wall tension. Hence, the $\pi$ Lagrangian does
not reproduce the relation between the deficit angle and the
tension \eqref{deficit} for the thin wall approximation.
Interestingly enough, it does reproduce the fact that, in order to
have a branch-changing 
solution, one has to provide a negative
energy density at some point.

\emph{Strong coupling}\\
If the branch-changing 
instanton existed, in principle there
would be the concern of how good an approximation to the
nucleation process it is, given that the brane would be at an
infinitely strongly coupled regime at the branch-changing 
point\footnote{The infinite strong coupling can be easily seen in
terms of the effective field theory \eqref{pi}. Consider for
simplicity that the only source is the brane tension $\sigma$. The
de Sitter invariant solutions are $\pi_0 = (c/2)\, x^\mu x_\mu $,
with $c=-K_\pm r_c\,\Lambda^3$ and $K_\pm$ given by \eqref{HRhoE},
and represent the de Sitter branes in the conventional or SA branch. The
fluctuations around this background can be studied by plugging
$\pi=\pi_0+\delta\pi$ in \eqref{pi}, and one obtains that the
coefficient in front of the kinetic term for $\delta \pi$ is $(1-2
K_\pm r_c)$. Hence, this mode becomes infinitely strongly coupled
for $K_\pm r_c=1/2$, which is precisely the point where the two
branches join, $\rho = - \rho_c$.}. One could expect that any
perturbation would produce a large backreaction, so that the
actual process significantly differs from the solution.

However, even if the infinitely strong coupling is achieved at the
branch-changing 
point, it may not prevent the solution from having
a physical meaning. The present situation is quite similar to the
violation of the WKB approximation in false vacuum decay. At the
turning points, the WKB approximation is broken maximally. However, in
the description of the bubble nucleation, the time coordinate is
extended from the real (or the imaginary) axis to the full complex
plane. Then the WKB approximation is violated only around 
a point,
e.g. $t=0$, in the complex plane. The situation concerning the strong
coupling seems to have the same structure in the complex
plane. Therefore having a region with strong
coupling would not render the analysis of instantons invalid.
This issue clearly deserves further investigation, though.

Let us emphasize that our main conclusion that
there is no evidence for the instability through
the branch-changing 
bubble nucleation is not affected by these
considerations. What we found is that there is no
branch-changing 
configuration even at the level of a
classical solution to the
Euclidean equations of motion.
In contrast, note that the DGP model does allow for solutions that
are infinitely strongly coupled on one time slice, although they
stay in the conventional branch at all times\footnote{ The
topology of the brane in such a solution is not $S^4$ but
$S^1\times S^3$ as long as a bubble wall composed of a scalar
field is concerned.}. Hence, in this sense, there is a clear
asymmetry between the two branches. This suggests that there is a
deep reason for it, and it might be tempting to associate it with
the presence of ghosts in the SA branch.\\

\begin{figure}[t]
\label{nullbubble}
 \begin{center}
\includegraphics[width=6cm]{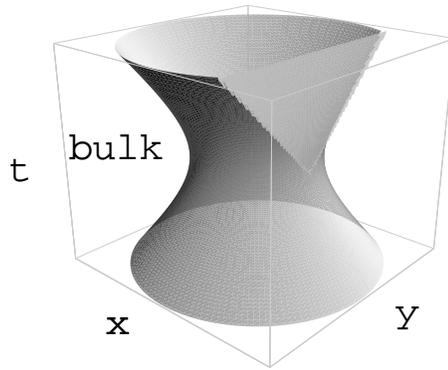}
\caption[]{Null bubble in the full DGP model. The two branches are
connected on a light cone. The upper half of this geometry is
given by the continuation of the instanton in Fig.~1 in the limit
when the bubble wall tension vanishes. The lower half is the
`background' solution, given by the SA branch. In the $\pi$
language, it corresponds to the solution \eqref{null}. }
\end{center}
\end{figure}

\emph{Zero radius bubble}\\
Another interesting solution is the ``zero-radius''
instanton\footnote{We thank Jaume Garriga for bringing our
attention to this limit.}. In Sec.~II we obtained an instanton
solution with a positive tension thin wall. Although we have shown
that this thin wall instanton cannot be understood as a physically
meaningful solution in Sec.~III, its zero-radius limit requires
more careful consideration. As the bubble radius shrinks, the
tension of the wall goes to zero. Hence, the zero-radius instanton
does not seem to require any matter field. In the Euclidean space,
the solution is a trivial sphere except for one point. However,
its analytic continuation to the Lorentzian region describes a
formation of a bubble of the conventional branch from a single
spacetime point. The junction of the two branches is given by a
null hypersurface as shown in Fig.~4. Although we call this
solution an instanton, the difference between the instanton action
and the background one vanishes. Therefore one can say that there
is no barrier for this process. Namely, the process could occur
classical mechanically if the solution turned out to be physical.

It is not clear whether this instanton configuration solves the
equations of motion in the DGP model or not. To answer this
question, we should carefully study the junction at the null
hypersurface\footnote{Again, it is possible to understand this
solution using the effective theory for the brane bending mode
$\pi$. The non-linear equation \eqref{pi} admits a solution of the
form
\begin{equation}\label{null}
\pi = -(\Lambda^3/2) \,x^\mu x_\mu\, (1- \Theta(-x^\mu
x_\mu)\Theta(x^0)),
\end{equation}
where $\Theta$ is a Heaviside step function. In the interior of
the future light cone, $x^\mu x_\mu <0$ and $x^0>0$, the solution
corresponds to the conventional branch and the remaining part is
in the SA branch. (Similar solutions with the bubble restricted to
$x^0>0$ can also be constructed in the presence of the brane
tension.) The only non-trivial part of \eqref{null} is what
happens at the light cone and at the origin. Besides the origin,
one can easily see that the localized distributional terms
appearing in the equation of motion are of the form
$\xi\delta(\xi)$, $\xi^2\delta'(\xi)$ and $(\xi\delta(\xi))^2$
where $\xi=(x^\mu x_\mu)^{1/2}$. While the first two can be taken
to be zero in a distributional sense, it is not completely clear
that one can set $(\xi\delta(\xi))^2$ to zero. However, if we use
a representation of the delta function defined as a limit of a
function, e.g. $\delta(\xi)\equiv\lim_{\epsilon\to
0}(\sqrt{2\pi}\epsilon) \exp(-\xi^2/2\epsilon^2)$, this term
vanishes. Up to this subtlety, this is a solution of the $\pi$
Lagrangian. }. However, this is not what we wish to address here.

In fact, the inverse process from the conventional branch to the
SA branch as illustrated in Fig.~5 also seems to exist similarly.
Hence, if this type of solutions is physical, the consistency of
the DGP model in the conventional branch is compromised. The point
that we want to stress is that, even if the zero-radius instanton
solves the equations of motion, it is far from clear if it 
has any physical meaning.

In contrast to the case of the strong coupling issue that we
discussed above, we would say that the zero-radius instanton is
clearly outside the range of validity of the present effective
theory at least at the classical level. The DGP model has a strong
coupling length scale below which one cannot trust this effective
theory. Here, the zero-radius instanton contains a jump in its
metric, and we know that this jump cannot be understood as a limit
of any smoothed configurations. Hence, the configuration of the
zero-radius instanton is clearly outside the range of validity of
the effective theory.

Furthermore, if we accept this solution as a physically meaningful
one, classical initial-value problems become non-deterministic.
Namely, the future evolution of a given initial data has various
possibilities\footnote{A similar situation was found
for `stealth'  branes \cite{stealth}.}. Such a situation should
not be allowed when we consider not a low energy effective theory
but a complete theory. In an appropriate UV completion for the DGP
model, the zero-radius instanton must not be actually a solution.
For this reason, we conclude that this zero-radius instanton is
excluded from the physical solutions at the classical level.

One may still think that this instanton has some meaning as a
quantum process. One is then lead to the logic that we used above
regarding the issue of the strong coupling for hypothetical smooth
instanton solutions. There, we extended the time coordinate to the
full complex plane. Once we do so, it may look unclear whether the
tunnelling process described by this instanton is really forbidden
or not. However, since the zero-radius instanton is not analytic,
containing a step function, its analytic continuation to the
complex valued time does not make sense. Therefore the
regularization method with the aid of the complex valued time does
not work for the zero-radius instanton. So, it is not clear if the
configurations in Fig.~4 and Fig.~5 have a physical meaning even
in this sense.

\begin{figure}[t]
\label{inversenullbubble}
 \begin{center}
\includegraphics[width=5cm]{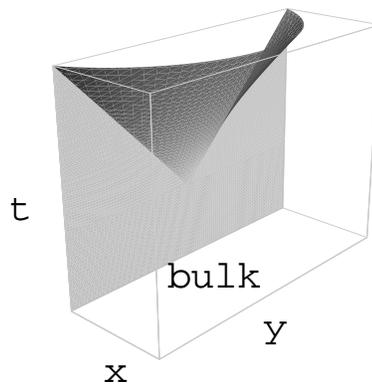}
\caption[]{Inverse null bubble in the DGP model. In this
configuration, the brane is in the SA branch in the interior of
the future light cone of some event.}
\end{center}
\end{figure}

\emph{Bubbles in the bulk}\\
Let us mention that another possible instability is the nucleation
of SA branes within the bulk, resembling bubbles of nothing. This
process was considered in \cite{kaloper}. The nucleated SA branes
would eat up the bulk and eventually collide with the other (our)
brane. This can happen irrespective of the branch in which our
brane is and for any value of its tension.
This suggests that this is not related to the perturbative ghost
instability, for the conventional branch does not have any ghost
mode. Furthermore, if the brane tension is negative, one can have
a similar instability due to the the nucleation of conventional
branch bubbles in the bulk.

In Ref.~\cite{kaloper}, the decay rate was estimated as $\sim
\exp(- r_c^2/\kappa_4^2)$ when the brane tension is small (a
similar instability was discussed in Ref.~\cite{stealth} when the
$Z_2$ symmetry across the branes is not imposed). Even though this
process is suppressed, it represents a catastrophic instability if
the light cone extends into the bulk forever in the past. However,
once we introduce a cutoff in time, typically of order the age of
the Universe $H_0^{-1}\sim r_c$, the probability that an observer
is hit by such a SA brane nucleated in the bulk becomes really
suppressed.
Let us just add that one can also avoid this instability trivially
by restricting to topology preserving configurations.\\

\emph{Taming the ghost?} \\ It is still unclear what is the
consequence of the ghost instability in the SA
universe. Usually, the ghost leads to spontaneous pair creation of
ghost and normal particles. Once such a channel opens, Lorentz
invariance leads to a divergence of the particle creation rates.
If we consider the situation in which there is a spin-2 ghost, we
need to treat the helicity zero mode in a different way.
Otherwise, negative norm states appear. But if we take a different
prescription for the quantization for the helicity zero mode, this
procedure necessarily breaks de Sitter invariance. When there is a
spin-0 ghost, a similar phenomenon happens. In this case, the mass
of the ghost is given by $-4H^2$. But we know that there is no de
Sitter invariant vacuum state for a scalar field with negative
mass squared. Once de Sitter invariance is broken, one may be
allowed to consider the possibility that the non-covariant cutoff
scale may arise due to the strong coupling effect. Then the
particle creation could be milder than the usual ghost in the
Minkowski background. Further studies are necessary to verify
this, but together with the non-existence of the non-perturbative
instability, the ghost in the SA universe could be
less harmful than expected.\\

\section*{Acknowledgments}

We would like to thank Gia Dvali, Gregory Gabadadze, Jaume
Garriga, Alan Guth, Nemanja Kaloper, Matt Kleban, Michele Redi and
Alex Vilenkin for useful discussions.
KK is supported by PPARC/STFC. OP acknowledges support from DURSI
(Generalitat de Catalunya), under contract 2005 BP-A 10131. TT is
partially supported by Monbukagakusho Grant-in-Aid
for Scientific Research Nos.~17340075 and~19540285. 
This work is also supported in part by the 21st Century COE
``Center for Diversity and Universality in Physics'' at Kyoto
 university, from the Ministry of Education,
Culture, Sports, Science and Technology of Japan,
and by the Japan-U.K. Research Cooperative Program
both from Japan Society for Promotion of Science.

\end{document}